\documentclass[aps,prd,10pt,twocolumn,preprintnumbers,superscriptaddress,bibnotes,nofootinbib,letterpaper,longbibliography]{revtex4-2}

\usepackage{hyperref}
\usepackage{graphicx}
\usepackage{amsfonts,amsmath,amssymb,bm}
\usepackage{physics}
\usepackage{svg}
\usepackage{array}
\usepackage{dcolumn}
\usepackage{multirow}
\usepackage{booktabs}

\newcolumntype{d}[1]{D{.}{.}{-1}}
\newcolumntype{C}[1]{>{\centering\let\newline\\\arraybackslash\hspace{0pt}}m{#1}}

\hypersetup{
  colorlinks  = true,
  urlcolor    = blue,
  linkcolor   = blue,
  citecolor   = red
}

\newcommand{\orcid}[1]
{\begingroup
  \hypersetup{hidelinks}\href{https://orcid.org/#1}{\includegraphics[width=9pt]{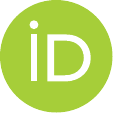}
} \endgroup}

\begin{document}

\title{Neutron Tagging Can Greatly Reduce Spallation Backgrounds in Super-Kamiokande}

\author{Obada Nairat \orcid{0000-0003-2019-9021}}
\email{nairat.2@osu.edu}
\affiliation{Center for Cosmology and AstroParticle Physics (CCAPP), \href{https://ror.org/00rs6vg23}{Ohio State University}, Columbus, OH 43210}
\affiliation{Department of Physics, \href{https://ror.org/00rs6vg23}{Ohio State University}, Columbus, OH 43210}

\author{John F. Beacom \orcid{0000-0002-0005-2631}}
\email{beacom.7@osu.edu}
\affiliation{Center for Cosmology and AstroParticle Physics (CCAPP), \href{https://ror.org/00rs6vg23}{Ohio State University}, Columbus, OH 43210}
\affiliation{Department of Physics, \href{https://ror.org/00rs6vg23}{Ohio State University}, Columbus, OH 43210}
\affiliation{Department of Astronomy, \href{https://ror.org/00rs6vg23}{Ohio State University}, Columbus, OH 43210} 

\author{Shirley Weishi Li \orcid{0000-0002-2157-8982}}
\email{shirley.li@uci.edu}
\affiliation{Department of Physics and Astronomy, \href{https://ror.org/04gyf1771}{University of California}, Irvine, CA 92697}

\date{\today}

\begin{abstract}
Super-Kamiokande's spallation backgrounds --- the delayed beta decays of nuclides following cosmic-ray muons --- are nearly all produced by the small fraction of muons with hadronic showers.  We show that these hadronic showers also produce neutrons; their captures can be detected with high efficiency due to the recent addition of dissolved gadolinium to Super-Kamiokande.  \textit{We show that new cuts based on the neutron tagging of showers could reduce spallation backgrounds by a factor of at least four beyond present cuts.}  With further work, this could lead to a near-elimination of detector backgrounds above about 6~MeV, which would significantly improve the sensitivity of Super-Kamiokande.  These findings heighten the importance of adding gadolinium to Hyper-Kamiokande, which is at a shallower depth.  Further, a similar approach could be used in other detectors, for example, the JUNO liquid-scintillator detector, which is also at a shallower depth.
\end{abstract}

\maketitle


\section{Introduction}
\label{sec:intro}

Crucial open questions in physics and astronomy depend upon improved measurements of solar neutrinos.  These questions include precision tests of neutrino mixing~\cite{ Palazzo:2011rj, Haxton:2012wfz,  Maltoni:2015kca, Capozzi:2017auw, Seo:2018rrb, Gann:2021ndb} and of the physical conditions in the solar interior~\cite{Asplund:2009fu, Turck-Chieze:2010rvs, Antonelli:2012qu, Vinyoles:2016djt, Gann:2021ndb}. Here we focus on neutrinos above a few~MeV, for which Super-Kamiokande (Super-K) has the largest statistics, due to operating since 1996 with a fiducial volume of 22.5~kton~\cite{Super-Kamiokande:2002weg, Abe:2013gga}.  In 2020, Super-K began adding dissolved gadolinium to enable neutron detection, as proposed in Ref.~\cite{Beacom:2003nk} and developed in Refs.~\cite{Marti:2019dof, Super-Kamiokande:2021the, Super-Kamiokande:2024kcb}.  While a major motivation for neutron tagging is improving sensitivity to the diffuse supernova neutrino background (DSNB)~\cite{Ando:2004hc,Beacom:2010kk, Lunardini:2010ab, Super-Kamiokande:2021jaq}, we show here that this new capability can also help solar neutrino (and other) studies by greatly reducing spallation backgrounds.  In Ref.~\cite{Beacom:2003nk}, neutron tagging was used to identify \textit{signals}; here we use it to identify \textit{backgrounds} through their pre-activity.  

A key component of the detector backgrounds is spallation, meaning nuclear breakup processes caused by cosmic-ray muons and their secondaries that lead to delayed beta decays.  In Super-K, these are dominant above $\sim$6~MeV, with beta decays due to intrinsic radioactivities dominating at lower energies~\cite{Super-Kamiokande:2015xra, Zhang:2016gxj, Super-Kamiokande:2021snn}.  Even though Super-K is shielded by 1~km of rock~\cite{Super-Kamiokande:2002weg, Abe:2013gga}, the rate of cosmic-ray muons that reach the detector is $\sim$2~s$^{-1}$~\cite{Super-Kamiokande:2005wtt} and the probability of a muon producing a relevant isotope is $\sim$0.02, corresponding to a spallation decay rate of $\sim$3000~day$^{-1}$ above 3.5~MeV.  In contrast, the signal rate from solar neutrino-electron scattering is $\sim$20 day$^{-1}$ above 3.5~MeV~\cite{Super-Kamiokande:2023jbt}.  Super-K has developed empirical background cuts, including one for approximately localizing isotope production along the muon track~\cite{Super-Kamiokande:2005wtt, Super-Kamiokande:2008ecj, Super-Kamiokande:2011lwo, Super-Kamiokande:2010tar, Super-Kamiokande:2016yck,Super-Kamiokande:2021snn}.  These cuts remove $\sim$90\% of spallation backgrounds at a cost of $\sim$10\% detector deadtime (signal loss).  Taking into account the directionality of the signal helps by another factor $\sim$10, but backgrounds comparable to the signal remain.  Therefore, there is sensitivity to be gained by further reducing spallation backgrounds, which requires a better physical understanding of their origins.

To date, the only detailed theoretical studies of spallation in Super-K were by Li and Beacom~\cite{Li:1, Li:2, Li:3}. In their first paper, they showed that nearly all relevant isotopes are made by muon secondaries, with their calculations reproducing Super-K's measured isotope yields within a factor of a few, consistent with expected hadronic uncertainties.  In their second paper, they showed that nearly all of these isotopes are produced in showers, which cause peaks in the muon Cherenkov light intensity along the track.  Further, they showed that while hadronic showers are rare compared to electromagnetic showers, they dominate isotope production.  In their third paper, they showed how to localize showers along the muon track much more precisely than done by Super-K.  As part of this work, Li and Beacom found that spallation isotopes are often produced in association with neutrons, with their yields increasing with shower energy (see also Ref.~\cite{Li:2016kra}).  They privately shared this idea with Super-K and encouraged its development. In Refs.~\cite{Super-Kamiokande:2021snn, Super-Kamiokande:2023jbt}, Super-K demonstrated an improved spallation cut based on localizing showers by requiring the production of clouds of neutrons.  Super-K's results so far are for pure water, in which neutron captures are detected with low efficiency, so their focus is on large showers, which are rare.

In this paper, we present the first detailed theoretical calculation of neutron production in association with spallation isotopes.  We focus on both the underlying physics and on new cuts for Super-K with added gadolinium.  We show that the majority of relevant isotopes are produced by a minority of showers that also produce neutrons.  If these dangerous showers (hadronic or electromagnetic with a hadronic or neutronic component) can be identified, they can be subjected to stronger cuts without incurring significant deadtime. \textit{In terms of the underlying physics, it is enough to tag showers with as few as one neutron, allowing sensitivity to even small showers.  Added gadolinium makes this feasible.}

The ability to use neutron tagging to identify potential isotope-producing showers depends on the neutron detection efficiency.  We focus on the present Super-K with gadolinium, showing that it can greatly reduce spallation backgrounds relative to the previous Super-K with pure water.  This highlights the importance of considering added gadolinium for Hyper-Kamiokande (Hyper-K)~\cite{Abe:2011ts, Hyper-Kamiokande:2018ofw}, which will be at a shallower depth, where the flux of muons is about five times higher~\cite{Hyper-Kamiokande:2018ofw}.  Further, the physical insight that isotope production often happens in hadronic showers with multiple neutrons will also be important for scintillator detectors like JUNO~\cite{JUNO:2015zny, JUNO:2015sjr, JUNO:2021vlw}, which is also at a shallower depth. \textit {As detailed in the Appendices, our insights significantly improve upon a standard technique (TFC, or three-fold coincidence between muon, neutron, and candidate isotope) used to reject $^{11}$C in scintillator detectors ~\cite{Galbiati:2004wx, Borexino:2011cjz}. It also generalizes upon new approaches used by Borexino and KamLAND-Zen that incorporate neutron multiplicity to tag $^{11}$C and xenon spallation events~\cite{Borexino:2021pyz, KamLAND-Zen:2023spw}.}

This paper is organized as follows. In Sec. II, we review the physics of particle showers and their production of isotopes. In Sec. III, we calculate the neutron yields of showers induced by hadrons and electrons.  In Sec. IV, we present our main results: we predict the neutron yields for muons in Super-K,  show that the vast majority of isotope-producing muons also produce neutrons, and propose a technique to utilize neutron yields as a method to tag those muons.  Finally, in Sec. V, we conclude.  In Appendices, we provide further details, including some new results.


\section{Review of spallation isotope production in Super-K}
\label{sec:Review}

In this section, we briefly review the physics of cosmic-ray muons in Super-K, their propagation through the detector, how they produce showers, how those showers produce isotopes, cuts to reduce the relevant isotopes, and neutron detection.


\subsection{Exposure to cosmic-ray muons}
\label{subsec:MuonsExposure}

Super-K is a water-Cherenkov neutrino detector built 1~km underground in the Kamioka mine in Japan~\cite{Super-Kamiokande:2002weg, Abe:2013gga}. The detector consists of a cylindrical stainless steel tank that is 41.4~m tall and 39.3~m in diameter, containing 50~kton of water. This volume is separated into two concentric cylinders, with the densely instrumented inner detector being 36.2~m in height and 33.8~m in diameter, and the remaining volume being the optically isolated and sparsely instrumented outer detector (and a 0.5-m dead zone between the inner and outer detectors). The fiducial volume is a virtual cylinder with each side 2~m inside the inner detector, making it about 32.2~m in height and 29.8~m in diameter, holding 22.5 kton of water.

Despite the rock overburden~\cite{Super-Kamiokande:2002weg, Abe:2013gga}, Super-K is exposed to cosmic-ray muons at a rate of $\sim$2 s$^{-1}$~\cite{Super-Kamiokande:2005wtt}.  At the depth of Super-K, the muons have an average energy of 271~GeV (almost all throughgoing), with most of the distribution being in the energy range 30--700~GeV~\cite{Tang:2006uu}.  Isotope production rates are only moderately sensitive to the detailed shape of Super-K's muon spectrum.  Most muons are relatively downgoing, with their angular distribution peaking at vertical and with $\sim$80\% of them being within $60^{\circ}$ of vertical. Taking into account the variations in the muon entry points and angles, the average muon track length inside the fiducial volume is $\sim$24~m~\cite{Super-Kamiokande:2022cvw} (for comparison, the longest diagonal path is $\sim$44~m). The majority ($\sim$89\%) of muons arrive as individual throughgoing particles with long pathlengths, but $\sim$7\% of them arrive in bundles with parallel tracks due to being generated by the same primary cosmic ray (described further in Sec.~\ref{subsec:Cuts}), and $\sim$4\% are corner-clipping muons with small track lengths, usually less than $\sim$7~m. 

In the following, as in Refs.~\cite{Li:1, Li:2, Li:3}, we focus on single vertical throughgoing muons that enter the detector at its center, which is sufficiently general to illustrate our new physics ideas.  Super-K has excellent muon reconstruction capabilities due to independent measurements in the inner and outer detectors, so different event classes are easily identifiable.  Further, isotope yields are calculated per length of (single) muon track (in units of $\mu^{-1} \, \text{g}^{-1}\, \text{cm}^2$), so the yields for varying entry points and angles can be rescaled by the muon track lengths.  Muon bundles are subjected to strong cuts. Corner-clipping muons can also be subjected to strong cuts; in any case, they have small track lengths, so their contribution to isotope production is small. We also do not consider low-energy muons that stop inside the detector ($\sim$5\% of all muons), which typically have small isotope yields except due to the nuclear captures of negative muons.  This case, considered in detail in Refs.~\cite{Super-Kamiokande:2000kzn, Li:2}, can be dealt with by excluding small spherical regions around the muon stopping points, which induces minimal deadtime~\cite{Super-Kamiokande:2005wtt}.

We simulate muon propagation in Super-K using the particle transport code \texttt{FLUKA} (version 4-4.0)~\cite{Ferrari:2005zk, Battistoni:2007zzb, Bohlen:2014buj}, which is a Monte Carlo simulation package that has been previously used to reproduce measured spallation data for a variety of detectors within hadronic uncertainties.  Other information also suggests that \texttt{FLUKA} models showers well~\cite{Borexino:2013cke, Locke:2020kco, Coffani:2021gbe, Super-Kamiokande:2021snn}.  We use a very similar setup to those detailed in Refs.~\cite{Li:1, Li:2, Li:3, Zhu:2018rwc}.  Our simulations are restricted to the fiducial volume of Super-K, with the target material being the natural isotopic abundance of water as predefined by \texttt{FLUKA}, though the effects of $^{17}$O and $^{18}$O are negligible~\cite{Li:1}. All relevant hadronic and electromagnetic processes are switched on using the PRECISIOn defaults. The isotope yields and neutron yields are calculated using a modified user routine that counts both relevant isotopes (without a 3.5~MeV threshold cut as it makes little difference~\cite{Li:1}) and neutron capture signals. Isotope yields from our modified user routine are also cross-checked with the built-in RESNUCLEi card. The neutron yields are calculated assuming Super-K's present gadolinium concentration (0.033\% by mass).  Gadolinium only affects neutrons once they are slowed to thermal energies, after which their travel distances are small without or with gadolinium. And, while we include gadolinium in our simulations, it is irrelevant as a spallation target due to its small concentration.

Even if the measured isotope yields differ from predictions, due to the hadronic uncertainties, that is acceptable.  First, we need only reasonable accuracy to design cuts that can be studied in further detail through Super-K measurements and simulations.  Second, the experimental unknowns are the spallation \textit{yields}, as the \textit{energy spectra} and \textit{time profiles} of almost all of the relevant isotopes are known from laboratory measurements.  Simulations could thus be recalibrated with measured yields.


\subsection{Muon propagation in water}

As muons propagate through water, they lose energy at a rate~\cite{Dutta:2000hh, Groom:2001kq, ParticleDataGroup:2024cfk}
\begin{equation}
\label{eq:MuonLoss}
-\frac{dE}{dX} = \alpha(E) + \beta(E) E,
\end{equation}
where $X$ is the mass column density ($\rho L$, the product of the mass density and the path length, which is quoted in units of g cm$^{-2}$), and where $\alpha(E)$ and $\beta(E)$ are the ionization and radiation coefficients, respectively. The energy dependence of $\alpha(E)$ and $\beta(E)$ is modest; the majority of the energy dependence in Eq.~(\ref{eq:MuonLoss}) arises from the explicit $E$ factor in the second term.  The muon critical energy, $E_c$, is the energy at which the ionization and radiation losses are equal. In water, this is 1.03~TeV~\cite{Groom:2001kq, Desai:2004mz}.  At the average energy of 271~GeV, ionization losses are dominant but radiation losses are appreciable (and cause the most relevant showers).

The $\alpha(E)$ term accounts for the energy losses due to the ionization or excitation of bound atomic electrons~\cite{Groom:2001kq, Bichsel:2006cs, ParticleDataGroup:2024cfk}. At 271~GeV in water, these losses have an average rate of 2.9~MeV~g$^{-1}$~cm$^2$~\cite{Groom:2001kq}. This can be separated into two components based on the energy transferred to the electrons. One component (the restricted ionization loss) leads to low-energy scattered electrons and is mostly continuous along the muon path, having only small fluctuations. Muons in Super-K have an average restricted ionization loss rate of $\sim$2~MeV~g$^{-1}$~cm$^2$. Another component (the delta-ray loss) can create high-energy electrons that can induce electromagnetic showers. Delta-ray losses, which have an average rate of $\sim$0.9~MeV~g$^{-1}$~cm$^2$, have large fluctuations and can produce delta rays with energies approaching that of the muon itself~\cite{Groom:2001kq, ParticleDataGroup:2024cfk}.

\begin{figure}[t]
\includegraphics[width=\columnwidth]{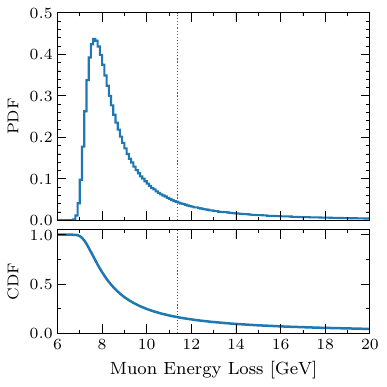}
\caption{Top panel: Differential distribution for the energy loss of muons passing vertically through Super-K. Bottom panel: Cumulative distribution for the fraction of muons with an energy loss above a given value. The red dotted line shows the average muon energy loss.}
\label{fig:MuonLoss}
\end{figure}

The $\beta(E)$ term accounts for radiation energy losses through interactions with atomic nuclei. The most relevant processes here are pair production, bremsstrahlung, and photonuclear interactions. Pair-production losses are almost continuous (with small fluctuations), with an average rate of $\sim$0.4~MeV~g$^{-1}$~cm$^2$, so they contribute to the minimum energy loss of muons crossing the detector.  The other processes have much larger fluctuations, often yielding energetic daughter particles that can, in turn, initiate electromagnetic or hadronic showers~\cite{Desai:2004mz}.  \textit{Importantly, the dominant energy losses of muons are due to electromagnetic processes, so electromagnetic showers are common and hadronic showers are rare.}

Figure~\ref{fig:MuonLoss} shows the energy loss distribution for vertical thoroughgoing muons in Super-K, which have a path length of 32.2~m, corresponding to an average restricted ionization loss of $\sim$6.4~GeV and an average pair-production loss of $\sim$1.3~GeV. These two losses are nearly continuous (with small fluctuations), causing an energy loss of $\sim$7.7~GeV on average, which is very close to the most probable muon energy loss at 7.6~GeV in the distribution. The tail to the left of the peak is due to the small fluctuations in these processes, which causes the minimum ionization loss to fluctuate down to $\sim$6.7~GeV in some cases. The average total energy loss is $\sim$11~GeV (more precisely, 11.4~GeV, as marked in Fig.~\ref{fig:MuonLoss}), which shows the importance of the tail caused by the highly fluctuating delta-ray and radiation losses.  We call the energy loss beyond the minimum the shower energy, as it is due to delta-ray production, photonuclear interactions, and other radiation processes that can produce showers if the secondary particles are energetic enough.  On average, the shower energy is $\sim$4~GeV, but fluctuations allow it to be much larger.  We caution that this shower energy is not perfectly defined because even what we call the minimum energy loss has some fluctuations.


\subsection{Showers induced by muons}

In a shower, an initial particle produces more and more particles of lower and lower energy; this multiplication ends when the average particle energy becomes low enough that the energy-loss or particle-loss rate is greater than the particle-production rate.  The energy where this occurs is the critical energy, $E_c$, already noted in the context of Eq.~(\ref{fig:MuonLoss}).

In electromagnetic showers, the secondary particles are mainly electrons, positrons, and gamma rays. The main processes driving the evolution of the particle shower are pair production and bremsstrahlung. As the shower develops, an electron or positron undergoes a bremsstrahlung interaction that produces an accompanying gamma ray. In turn, a gamma ray undergoes pair production, yielding an electron and a positron. Occasionally, secondary gamma rays in electromagnetic showers undergo photonuclear interactions that produce neutrons, and less often charged pions that, in turn, can develop a hadronic component. This continues until the critical energy is reached; for electromagnetic showers in water, this is about 80~MeV~\cite{ParticleDataGroup:2024cfk}. At that average energy, the number of particles in the shower reaches a maximum. 

Hadronic showers develop similarly to electromagnetic showers, but the main secondary particles are charged ($\pi^+,\pi^-$) and neutral ($\pi^0$) pions. Successive inelastic hadronic interactions lead to particle multiplication in each generation. The charged pions generally go through a succession of inelastic collisions, leading to nuclear breakup and the production of pions, until all hadrons fall below the critical energy.  In water, this is $\sim$1~GeV; this high critical energy is part of why hadronic showers are less common than electromagnetic showers.  The neutral pions promptly decay to gamma rays, $\pi^0 \rightarrow \gamma + \gamma$, which in turn can initiate electromagnetic showers. Since the pions are produced in hadronic showers with roughly equal numbers of each charge, about $1/3$ of each generation's energy is transferred to the electromagnetic shower (with an ultimate ceiling of about 90\% of the initial shower energy~\cite{Beatty:2009zz, Rott:2012qb}).  This electromagnetic component of a hadronic shower thus dominates its total light yield, with that being somewhat reduced compared to a pure electromagnetic shower. This is because charged pions have higher Cherenkov thresholds than electrons and because they lose energy to neutral particles (especially neutrons) that themselves produce no Cherenkov light~\cite{Li:2}.

The length scale of particle showers is determined by the mean free path of their underlying processes in the medium they transverse~\cite{Greisen:1960wc, Lipari:2008td, ParticleDataGroup:2024cfk}. For electromagnetic showers in water, the mean free path of bremsstrahlung and pair production processes, or the radiation length $X_0$, is about 36~cm~\cite{ParticleDataGroup:2024cfk}.  The distance to the shower's maximum is $X_0 \ln(E_0/E_c)$, where $E_0$ is the initial energy of the first particle starting the shower.  The full length of the shower is several times larger~\cite{Heitler:1936jqw, Matthews:2005sd}.  For hadronic showers, the mean free path for pion interactions is numerically somewhat larger than $X_0$, but the multiplicity of the particles in each generation is larger than in electromagnetic showers~\cite{Matthews:2005sd}, leading to a comparable length for the showers. In Super-K, muon-induced electromagnetic and hadronic showers are less than 10 meters long even at the highest relevant energies, so they are often fully contained within the fiducial volume (when they are not, they can be treated as showers of lower energy).

Electromagnetic showers are quite narrow, with a characteristic (Moli\`ere) radius of $\sim$0.1 m in water~\cite{ParticleDataGroup:2024cfk}, while hadronic showers, which have larger fluctuations at a given energy, have a characteristic radius of $\sim$1~m, about twice the hadronic interaction length (see the Appendices for details).  While many of the secondary particles in both types of showers are collimated in the direction of the muon, lower-energy secondaries have a broader range of directions, which is important for shower reconstruction~\cite{Li:3}.

At energies below $\sim$300~MeV, where pion production is kinematically suppressed, another kind of hadronic shower --- sometimes called a nucleon cascade or neutronic shower~\cite{Peng:1943hh, Rossi:1952kt, Li:2} --- is possible.  In these showers, gamma rays can eject one or more nucleons from a nucleus.  Protons are quickly stopped in the medium by their electromagnetic energy losses, but neutrons can eject nucleons from other nuclei, often leading to more gamma-ray emission from nuclear de-excitation.  These showers are near isotropic and have an extent below $\sim$1~m.  Ultimately, after losing energy through inelastic and elastic scattering, the neutrons typically undergo radiative capture (in Super-K, on hydrogen unless gadolinium is present).


\subsection{Production of isotopes by showers}
\label{subsec:IsotopeProduction}

The vast majority of relevant isotopes are produced by secondary particles in showers, not the parent muons. The isotope yields can be calculated using the convolution of the path length distributions (the total distances covered by all particles of a given type and energy) with the cross sections of the various isotope-production processes.  This convolution can be done by hand (see, e.g., Refs.~\cite{Galbiati:2004wx, Galbiati:2005ft}) but is typically done internally by codes like \texttt{FLUKA}.

The total path length of the secondary particles in a shower is directly related to its initial energy because the number of secondary particles in a shower grows linearly with shower energy.  The total energy in showers can be estimated from the total amount of Cherenkov light produced by the muon, minus the expectation from the non-showering energy loss components for the muon~\cite{Super-Kamiokande:2007uxr, Radel:2012kw, Li:3}. Therefore, the larger the light yield of a shower, the greater the cumulative path length of its secondary particles, which leads to a higher probability of producing an isotope.

Many of the inelastic nuclear cross sections have been measured in laboratory experiments~\cite{NNDC}. For the most relevant isotopes, the primary processes of production are through neutrons and pions breaking oxygen nuclei~\cite{Li:1}. The abundance of both neutrons and pions in hadronic showers makes them very efficient at producing those isotopes.  (To give a sense of the multiplicities, at the end of a hadronic shower, the number of remaining neutrons is about ten times greater than the number of pions; we give details in Sec.~\ref{subsec:NeutronYields} and Appendix~\ref{appendix:Yields}.)  Next most important are gamma rays, which sometimes cause photonuclear interactions.  While the photonuclear cross sections are small, the high frequency of electromagnetic showers partially overcomes this.

Most of the isotopes produced by muon spallation in water do not lead to detector backgrounds, because the isotopes are stable, have very long half-lives, or the decays do not produce high-energy betas~\cite{Li:1}.  However, for a small fraction of unstable isotopes, their beta decays cause significant backgrounds for neutrino signals in the energy range of 5--20~MeV.  These relevant isotopes are sometimes called ``background isotopes;" unless we say otherwise, we always mean these when we simply say ``isotopes."  As an example, $^{16}$N is important due to its large yield ($\sim 20 \times 10^{-7} \, \mu^{-1} \, \text{g}^{-1}\, \text{cm}^2$), half-life ($\sim$7 s), and $Q$-value ($\sim$10.4~MeV)~\cite{Super-Kamiokande:2000kzn, Galbiati:2005ft, Super-Kamiokande:2005wtt, Li:1}.  For throughgoing muons, the production of $^{16}$N is dominated by $(n,p)$ reactions of secondary neutrons with oxygen nuclei. This makes it important to study the production of neutrons and other secondary particles in muon-induced showers to develop cuts on these isotopes.  (There is also an important contribution due to the nuclear capture of stopping muons~\cite{Galbiati:2005ft, Super-Kamiokande:2005wtt}.)


\subsection{Super-K cuts on spallation isotopes}
\label{subsec:Cuts}

Super-K has developed several cuts to reduce backgrounds caused by spallation. A simple way to understand those cuts is to think of them as excluding all particles in cylinders around the muon tracks for a certain amount of time after the muon.  For a simple estimate, cutting all events within cylinders of radius 1~m for 20~s can reach a rejection efficiency of $\sim$80\%. However, such cuts would cause a significant deadtime of $\sim$20\%~\cite{Li:1, Li:2}, which makes it important to develop better cuts.  (For these and related estimates, we consider only vertical downgoing muons and ignore overlaps of the cylinders; this is a good --- in fact, conservative --- approximation because the deadtime is small.)

Super-K's first improvement to this was developing a sophisticated log-likelihood function that takes into account three factors: the time difference between the signal and the preceding muon ($t$), the transverse distance between the signal and the muon track ($L_{\text{trans}}$), and the radiative energy loss (shower energy) of the muon ($Q_{\text{res}}$). As discussed above, the higher the energy of the shower, the more secondary particles it produces, and therefore the more likely it is to produce an isotope. Employing this likelihood cut along with spherical cuts around stopping muons, Super-K rejected $\sim$90\% of the background at a cost of $\sim$20\% deadtime~\cite{Ishino:1999xt, Super-Kamiokande:2001ljr, Super-Kamiokande:2005wtt, Super-Kamiokande:2008ecj, Super-Kamiokande:2010tar, Super-Kamiokande:2016yck}. 

Super-K's next improvement was in a DSNB analysis, when they discovered that the spallation decays were located near the peak in the Cherenkov muon light profile~\cite{Super-Kamiokande:2011lwo}.  However, the localization was poor and the underlying physics was not understood.  In Refs.~\cite{Li:2, Li:3}, these problems were solved.  The light yield of a muon arises from the muon itself plus the secondary particles produced in the induced showers that move along with the muon.  A peak in the muon light profile corresponds to the location of a muon-induced shower.  At this point, the number of secondary particles and the probability of producing an isotope are both at a maximum.  This realization was taken into account in Refs.~\cite{Super-Kamiokande:2011lwo, Li:3, Super-Kamiokande:2021snn} by adding another variable to the likelihood: the longitudinal distance ($L_{\text{long}}$) between the possible spallation decay and that peak's location. This can be thought of as a way to make the cylinders shorter in length, such that they only cover the shower region instead of the full muon track, reducing deadtime.

This improved likelihood cut was used in the latest solar neutrino analysis by Super-K~\cite{Super-Kamiokande:2021snn}. In addition to that, they used a preselection cut that targets clusters of two or more isotopes produced by the same muon, which they referred to as a multiple spallation cut (Borexino used a similar multiple-isoptope approach to tag $^{11}$C isotopes~\cite{Borexino:2021pyz}). Unlike the likelihood cut, the multiple spallation cut does not require muon reconstruction, but it does require the reconstruction of multiple isotope decays.  Overall, the latest Super-K analysis achieved an efficiency of $\sim$90\% with a 10.5\% deadtime~\cite{Super-Kamiokande:2023jbt}, a significant improvement over earlier work by Super-K. For the data period with neutron tagging in pure water, they also applied cuts based on clouds of neutrons (two or more detected, which implies $\sim$10 times more produced)~\cite{Locke:2020kco, Coffani:2021gbe, Super-Kamiokande:2021snn}.  These cuts used geometric estimates of the shower size and position based on the multiplicity and locations of the neutron captures.  The impact of these geometric cuts was limited by the low neutron statistics.  We advise that it is better to estimate the shower size and position based on its abundant Cherenkov light.  This neutron cloud technique further reduced the deadtime to 8.6\%~\cite{Super-Kamiokande:2021snn, Super-Kamiokande:2023jbt}.

As an aside, we advocate that the multiple spallation cut be replaced by a strong cut on muons with large total energy losses, as suggested in Ref.~\cite{Li:2}, which notes that muons with energy losses of 30~GeV or more produce $\simeq$60\% of isotopes and can be strongly cut while incurring only minimal deadtime. This approach has several advantages.  First, it does not require multiple isotopes, so it applies to lower-energy showers.  Second, it can be adapted to muons that are not vertically downgoing by using a cut at a maximum energy loss of $30~{\rm GeV}/32.2~{\rm m} = 0.93~{\rm GeV/m}$.  Third, as described next, it can be used effectively against large muon bundles.  Full reconstruction of the muon(s) should not be necessary; it should be enough to localize the approximate entry and exit positions in the outer detector and to measure the total light produced.

Muon bundles are dangerous if $N$ muons are assumed to be $N - 1$ or fewer, leading to a miscalculation of the distance of an isotope candidate from a muon track.  Based on MACRO data~\cite{MACRO:1992sry, MACRO:1992boa},  double muons should occur at the $\sim$10\% level and higher-multiplicity muons at the $\sim$1\% level; the typical separation between muons is several meters.  At present, Super-K attempts to reconstruct the positions and directions of simultaneous muons with even large multiplicities (up to 10~muons)~\cite{Conner:1997xj}.  For spallation work, this approach can be replaced with a simple cut based on the total energy.  For example, four vertically downgoing muons would deposit a minimum total energy of $\sim$27~GeV, close to the value noted just above; cutting these events based on energy alone (or generalizing to energy loss per meter) would greatly simplify the work needed.  Following reconstructions of not just the muon positions and directions, but also their energy-loss profiles~\cite{Li:3}, we expect that most of the cases with multiplicities of three muons could be distinguished from single showering muons and thus rejected.  For the most relevant case of two muons, those with large separations can be treated independently, while those with small separations can be treated as a single, brighter muon.


\subsection{Neutron detection in Super-K}
\label{subsec:NeutronDetection}

Once neutrons have downscattered to thermal energies~\cite{Fermi, Segre}, they undergo radiative capture.  In pure water, this takes place with a capture time of $\sim$200~$\mu$s, and proceeds as:
\begin{equation}
\label{eq:HCapture}
n \, + \,^1_1\text{H} \, \rightarrow \, ^2_1\text{H} \,+\,\gamma \,\,\text{(2.2~MeV)}
\end{equation}
For the Super-K Stage-IV data, the trigger efficiency of 2.2~MeV gamma rays within the fiducial volume is 17\%~\cite{Super-Kamiokande:2013ufi, Super-Kamiokande:2021snn}.  With an increasing concentration of dissolved gadolinium in the water, the neutrons will more likely capture on gadolinium, due to its enormous cross section.  At the present concentration of 0.033\% by mass, 75\% of neutrons capture on gadolinium, with a capture time of $\sim$60~$\mu$s~\cite{Super-Kamiokande:2024kcb}.  This proceeds as (for one example gadolinium isotope):
\begin{equation}
\label{eq:GdCapture}
n \, + \,^{157}_{\, \, \, 64}\text{Gd} \, \rightarrow \, ^{158}_{\, \, \,64}\text{Gd} \,+\, \Sigma\gamma \,\,(\sim\!\text{8~MeV}),
\end{equation}
where the $\sim$8~MeV is shared among a few gamma rays.  While detecting a 2.2-MeV gamma ray is very challenging, detecting a total of 8~MeV in gamma rays is much easier, as this has an electron-equivalent energy of 4--5~MeV~\cite{Super-Kamiokande:2019xnm}.  In Super-K, gamma rays are not detected directly, but only through the Cherenkov light produced by the relativistic electrons that they produce through Compton scattering (and, less often, pair production).

For Super-K with added gadolinium, we can neglect accidental coincidences of real or apparent neutron captures with muons.  As a simple estimate, the rate of \textit{all} 4--5~MeV events before spallation cuts is $\sim$$10^4$~day$^{-1}$ and the muon rate is $\sim$$10^5$~day$^{-1}$, so for a separation of $\sim$100~$\mu$s ($\sim$$10^{-9}$~day$^{-1}$), the accidental coincidence rate is $\sim$1~day$^{-1}$.  For Super-K with pure water, the very low energy threshold required to detect 2.2~MeV gamma rays opens the door to much larger backgrounds, such as intrinsic radioactivities and contributions from fake neutrons associated with dark noise and afterpulsing of the photomultiplier tubes.  These can make the rate of accidental coincidences appreciable, and Super-K uses various cuts to reject them.  For example, they cut on all events within $\sim$20~$\mu$s of a muon to account for afterpulsing, which causes them to miss the neutron captures that occur within that short timescale. Overall, these cuts reduce the tagging efficiency of neutrons in pure water to less than 10\%~\cite{Super-Kamiokande:2021snn}.  The (small) impact of fake neutrons is discussed in Sec.~\ref{subsec:NeutronsRare}.


\section{Neutron production in showers}
\label{sec:NeutronProductioninShowers}

In this section, we calculate the neutron yields of showers induced by hadrons and electrons as a function of their energy.  This is a prerequisite to making realistic calculations for Super-K based on their distributions of shower energies. Here, we define the neutron yields as the number of neutrons that remain at the end of a shower, reach thermal energies, and undergo radiative capture. As described in Sec.~\ref{subsec:NeutronDetection}, the timescales of the radiative captures are very long (in the $\mu$s range) compared to the shower development time (in the ns range). Note that this definition includes neutrons produced in the isotope formation itself through reactions like $^{16}$O($\pi^{-}$,$\alpha$+2p+n)$^{9}$Li, but not the neutrons produced in the eventual decays of some isotopes, like the $\beta^-n$ decay of $^9$Li. We also do not count neutrons that are absorbed during the shower, e.g., through $(n,p)$ interactions.


\subsection{Average neutron yields}
\label{subsec:NeutronYields}

\begin{figure}[t]
\includegraphics[width=\columnwidth]{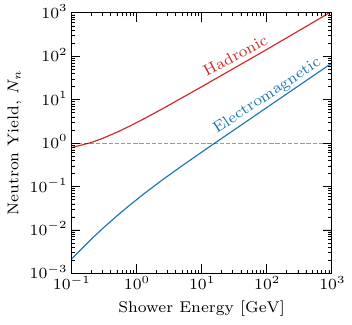}
\caption{Predicted neutron yields remaining at the end of hadronic and electromagnetic showers as a function of shower energy.  The dashed line indicates one neutron in the fiducial volume of Super-K.}
\label{fig:NeutronCount}
\end{figure}
\begin{figure*}[ht]
    \includegraphics[width=\columnwidth]{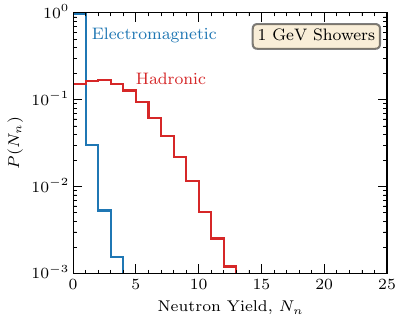}
    \includegraphics[width=\columnwidth]{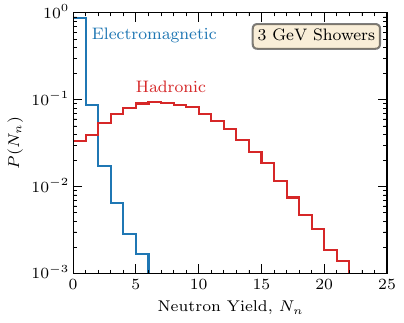}	
    \includegraphics[width=\columnwidth]{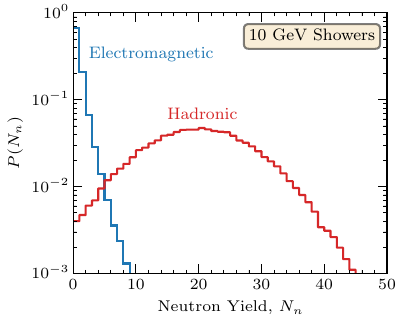}
    \includegraphics[width=\columnwidth]{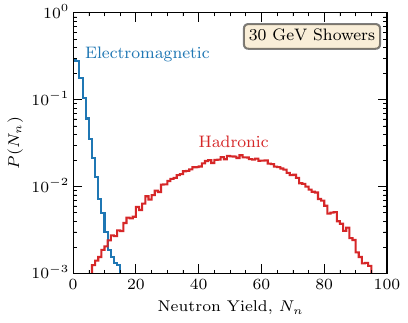}
\caption{Neutron yield distributions (bins of one count) for hadronic and electromagnetic showers (each normalized to unity, so the relative frequencies are not taken into account) of selected energies.  Note variations among the x-axes.  Here and in subsequent figures, we set the bin left edge (not the center) at the bin value, e.g., the bin for $N_n = 0$ runs from 0 to 1.}
\label{fig:NeutronProbabilityDistributions}
\end{figure*}

Neutrons, which are abundantly produced by muons in underground experiments, can cause a wide variety of difficult-to-reject detector backgrounds, not only through spallation decays.  This is why neutrons are sometimes called ``the rats of the lab."  There have been many measurements of neutron yields, but they have typically been reported as a function of the average energy of the muons at a given depth; see, e.g., Refs.~\cite{KamLAND:2009zwo, Borexino:2013cke, DayaBay:2017txw, Super-Kamiokande:2022cvw, RENO:2022xbr}. These measurements have found a direct relationship between the average muon energy and the average neutron (or spallation) yields. While this is useful, the average rate of neutrons does not fully capture the underlying physics due to the large fluctuations in the energy losses of individual muons, and hence their neutron yields. Instead, it would be better to measure the neutron yield as a function of the parent shower energy, as this would put measurements at different depths on a common basis closely related to the underlying physics.  As noted above, the shower energy can be determined from the measured total energy loss in the detector minus the predicted minimum energy loss.  (At energies below the critical energy, it would be more accurate to say injected energy, as proper showers will not develop.)  The neutron yield measurements would be even more informative if they were measured separately for hadronic and electromagnetic showers.

We simulate hadronic showers by injecting charged pions (equal fractions of $\pi^+$ and $\pi^-$) and electromagnetic showers by injecting electrons (or positrons or gamma rays).  On average, hadronic showers produce about ten times more neutrons than electromagnetic showers of the same energy.  As we show below, this difference is significantly larger than the expected fluctuations. Therefore, it should be easy to separate energetic hadronic and electromagnetic showers through measurements of the shower energy ($E_{\rm sh}$) and the neutron yield ($N_n$), bearing in mind that hadronic showers have somewhat lower light yields than electromagnetic showers for the same injected kinetic energy~\cite{Li:2, Li:3}.  The only difficulties would arise when $N_n \lesssim 1$.

Figure~\ref{fig:NeutronCount} shows the average neutron yields of hadronic and electromagnetic showers of different energies, where the shower energy is defined by the kinetic energy of the injected particle.  (One might instead use the total energy, as the mass energy of the pion is ultimately released through decay, but at the low energies where that distinction could matter, the particles do not actually shower.)  

In hadronic showers, neutrons are a major component.  When charged pions multiply through inelastic interactions with nuclei, they very often produce neutrons. The neutrons subsequently undergo their own inelastic collisions with nuclei, ejecting more neutrons (and protons, which quickly come to rest).  Below several~MeV, neutrons undergo elastic collisions until they reach thermal energies, where their capture cross section is largest, and eventually capture on a nucleus (in Super-Kamiokande, this is always hydrogen unless gadolinium is present).  At energies above $\sim$1~GeV, we find that $N_n$ rises as $\sim$$E_{\rm sh}^{0.9}$, where the slope is less than unity because of the energy lost to the electromagnetic component of the hadronic showers through neutral pion production.  At the lowest energies, the slope is flatter because injected pions have kinetic energies too low to induce a shower.  Injected $\pi^+$ simply decay to $\mu^+$, producing no neutrons, while injected $\pi^-$ capture on nuclei, ejecting one or more neutrons~\cite{Ponomarev:1973ya, Measday:2001yr}.

In electromagnetic showers, neutrons are a minor component. They arise from occasional photonuclear interactions, which typically produce neutrons that in turn induce neutronic showers when energetic enough. Less frequently, these interactions can generate charged pions, which can initiate small hadronic showers. At energies above $\sim$1~GeV, we find that $N_n$ rises as $\sim$$E_{\rm sh}$.  At lower energies, the slope is steeper because of a drop in the photonuclear cross section and because any produced pions are not energetic enough to induce a significant hadronic or neutronic component.

A key point is that a $\sim$1~GeV hadronic shower yields three neutrons on average.  This is around the critical energy of hadronic showers in water, which means that almost any proper hadronic shower induced by a muon is very likely to produce at least one neutron, even after downward statistical fluctuations. Meanwhile, electromagnetic showers must be above $\sim$20~GeV to produce even one neutron on average.  In subsequent sections, we show that these points can be used to reject showers that are likely hadronic (or have a hadronic or neutronic component) and thus likely produce isotopes.  Because electromagnetic showers have such a high frequency, they are moderately important for producing isotopes despite the small probability of their developing a hadronic or neutronic component.

In Appendix~\ref{appendix:Yields}, we give further details on the correlated yields of final-state neutrons, pions (and hence muons), and spallation isotopes, showing that they are all significantly higher for hadronic shower than electromagnetic showers (Fig.~\ref{fig:Yields}).


\subsection{Neutron yield distributions}
\label{subsec:NeutronYieldDistributions}

Figure~\ref{fig:NeutronProbabilityDistributions} shows our calculated neutron yield probability distributions for hadronic and electromagnetic showers at four specific energies.  These and similar distributions would allow the calculation of the probability that a given shower is hadronic or electromagnetic from its energy and neutron yield. The very small overlapping regions between these distributions (note the logarithmic scale on the y-axes) would allow determining the type of shower with great confidence.

As a simple example, for a 10-GeV shower, there is a crossover at about five neutrons.  Below this, a shower is much more likely to be electromagnetic; above it, it is much more likely to be hadronic.  However, the actual case is more complicated than this, because muon-induced showers in water are dominantly electromagnetic, therefore, an accurate classification requires taking into account the relative frequencies of muon-induced hadronic and electromagnetic showers.

In the next section, we show that a simpler approach is to use neutron yields to tag the most dangerous muon-induced showers, without the need to classify them as hadronic or electromagnetic. In other words, we can classify muons simply based on their neutron yields.


\section{Tagging spallation production with neutrons}
\label{sec:NeutronTagging}

In this section, we present our main results.  We calculate the spectra of muon-induced showers, the neutron yields of those showers, the association of isotopes with neutrons, the frequency of neutron-producing showers, and recommendations on how to implement our results. \textit{A key point is that while it would be helpful to separate showers as dominantly hadronic or electromagnetic, it is enough to separate them as neutron-producing or not.}  This then includes the small fraction of electromagnetic showers that develop hadronic or neutronic components and thus are also likely to produce isotopes.

As noted in Sec.~\ref{sec:intro}, we seek only to \textit{identify} a high probability of spallation production; a small number of neutrons is enough for this.  In contrast, Ref.~\cite{Super-Kamiokande:2021snn} sought to \textit{localize} the site of spallation production, which requires a large number of neutrons.  \textit{Our approach is thus sensitive to lower-energy, more-frequent showers.}


\subsection{Energy spectra of muon-induced showers}

\begin{figure}[t]
\includegraphics[width=\columnwidth]{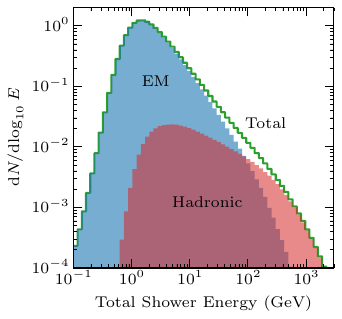}
\caption{Calculated energy spectra of muon-induced showers in Super-K, separated into electromagnetic and hadronic cases.  These are normalized per vertical throughgoing muon.}
\label{fig:ShowersEnergySpectrum}
\end{figure}

For the results that follow, we simulate showers produced by vertical throughgoing muons (using the spectrum from Ref.~\cite{Tang:2006uu}) passing through Super-K's fiducial volume.  This gives the energy spectra and types of muon-induced showers. To proceed, we must address three issues that were simple in Sec.~\ref{sec:NeutronProductioninShowers} but which have some ambiguities here.  These issues can all be dealt with reasonably.

\begin{enumerate}

\item 
Above, we call a shower hadronic if we initiated it with a charged pion and electromagnetic if we initiated it with an electron, positron, or gamma ray.  In practice, it is convenient to call a shower hadronic as long as it is partially so --- containing at least one pion or kaon --- which then includes electromagnetic showers that develop even a small hadronic component.  This approach helps identify showers that are likely to produce isotopes while keeping the frequency relative to the muon rate low enough.

\item
Above, we define the shower energy by the kinetic energy of the injected particle.  In practice, it is more convenient to define the shower energy as the total energy loss of the muon minus the minimum expected energy loss. While this definition of the shower energy is somewhat ambiguous due to fluctuations, this approach helps because it is easy to use and the exact energy scale is not important.

\item
Above, we define the shower energy assuming a single shower.  In practice, it is more convenient to sum the energy of all showers along a muon track.  It is common to have one significant shower be accompanied by smaller, unrelated showers along the same muon track.  This approach helps focus on the total probability of producing an isotope in association with a muon.

\end{enumerate}

Figure~\ref{fig:ShowersEnergySpectrum} shows the energy distribution of muon-induced showers in Super-K. This shows in detail the effects of the energy loss tail in Fig.~\ref{fig:MuonLoss}, following subtraction of the minimum energy loss, which we take to be 6.7~GeV.  Here we use a log scale on the $x$ axis to show the wide range of values.  Correspondingly, on the $y$ axis, we use $dN/d\log_{10}E = 2.3 \, E \, dN/dE$.  With this choice, the relative values at any two energies reflect their relative contributions to the integral.

We separate Fig.~\ref{fig:ShowersEnergySpectrum} into hadronic and electromagnetic showers, as defined in this subsection, which leaves the electromagnetic showers as being ``pure" (no pions), though they may still produce neutrons.  As we show below, with this definition, hadronic showers (rare) are effective at producing neutrons and isotopes while electromagnetic showers (common) are not.  This allows strong and effective cuts that focus on a minority of muons so that the efficiency and/or deadtime can be improved.

Electromagnetic showers dominate the spectrum, comprising $\sim$97\% of all showers in Super-K. This shower spectrum peaks at $\sim$1~GeV because the peak corresponding to the most probable energy loss in Fig.~\ref{fig:MuonLoss} is about 1~GeV above the minimum.  The tail at low energies reflects the small fluctuations in the continuous ionization and pair production losses.  The tail at high energies reflects the large fluctuations of the delta-ray, bremsstrahlung, and photonuclear energy losses, with the delta-ray component being the most important~\cite{Groom:2001kq, Li:2}.  For that process, the differential cross section scales with the energy transferred to electrons as $\sim \, 1/E^2$, so the plotted $EdN/dE$ spectrum falls as $\sim \, 1/E$.  At higher energies, the electromagnetic shower spectrum steepens because of the increasing probability of developing a hadronic component and thus being reclassified. 

Hadronic showers, which are subdominant, are mostly caused by photonuclear interactions of muons that produce charged pions, and less often from photonuclear interactions of gamma rays in electromagnetic showers. When any produced charged pions have energies above the hadronic critical energy of $\sim$1~GeV, they are likely to produce a shower with multiple generations of charged and neutral pions and a large neutron yield, making them very efficient at producing isotopes. Even when the pions are not energetic enough to induce showers, $\pi^-$ captures on nuclei often produce isotopes. 

The differential cross section of muon photonuclear interactions scales with the transferred energy to pions as $\sim \, 1/E^{1.5}$~\cite{Groom:2001kq}, which makes the hadronic spectrum less steep than that of electromagnetic showers. Pion-producing photonuclear interactions also have a higher kinematic threshold, causing the spectrum to be shifted to higher energies. The kinetic energy spectrum of individually produced pions has a wide distribution, with a slight peak near $\sim$0.4~GeV and a well-populated tail extending to much higher energies~\cite{Li:2}.  Muon photonuclear interactions often lead to the production of multiple pions, since we are summing the energy of all showers, the hadronic shower spectrum rises rapidly around a total shower energy of $\sim$1~GeV and peaks at $\sim$3~GeV. At the highest energies, there is a sharp drop because energy losses above $\sim$300~GeV require a muon with energy at least equal to that, where the spectrum of muons at Super-K falls as $1/E^3$ at these high energies~\cite{Tang:2006uu, Li:1}.
\subsection{Neutron yields of hadronic showers are large}

\begin{figure}[!t]
\includegraphics[width=\columnwidth]{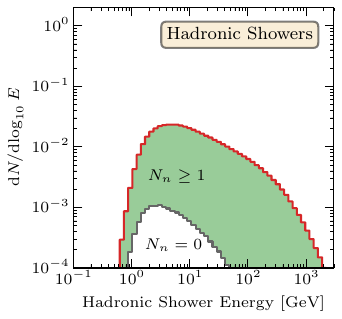}
\caption{Spectrum of muon-induced hadronic showers from Fig.~\ref{fig:ShowersEnergySpectrum}, separated into two components based on neutron yield, normalized per vertical throughgoing muon.}
\label{fig:HadronicNeutrons}
\end{figure}

\begin{figure}[t]
\includegraphics[width=\columnwidth]{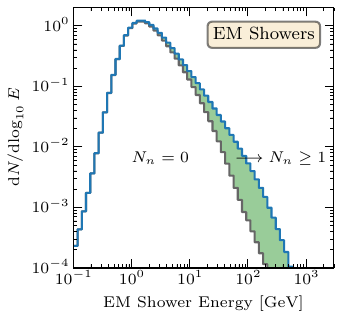}
\caption{Spectrum of muon-induced electromagnetic showers from Fig.~\ref{fig:ShowersEnergySpectrum}, separated into two components based on neutron yield, normalized per vertical throughgoing muon.}
\label{fig:emNeutrons}
\end{figure}

Here we calculate the neutron yields for different types of muon-induced showers in Super-K.

Figure~\ref{fig:HadronicNeutrons} shows that hadronic showers almost always produce neutrons.  As in Sec.~\ref{sec:NeutronProductioninShowers}, we define $N_n$ as the number of low-energy neutrons left at the end of a shower; this is the same number that will undergo capture, for which the detection efficiency depends on the gadolinium concentration.  The number of neutrons produced is often large (as shown above in Fig.~\ref{fig:NeutronProbabilityDistributions}).  If we remake Fig.~\ref{fig:HadronicNeutrons}, instead requiring, for example, $N_n \ge 3$, the results only change slightly.  The fraction of hadronic showers that do not produce any neutrons is only 3\%. The logarithmic scale on the y-axis highlights the differences --- which are statistically significant --- in the energy spectra between the $N_n=0$ and $N_n \ge 1$ cases. The majority of the showers without neutrons are either low-energy hadronic showers or electromagnetic showers with small hadronic components, both of which are less efficient at producing isotopes anyway.  These points mean that neutrons are extremely \textit{efficient}  at identifying hadronic showers, which is good.

Figure~\ref{fig:emNeutrons} shows that electromagnetic showers almost never produce neutrons, with the probability to obtain even one neutron being only 9\%.  As a reminder, we have defined a shower as hadronic if it has even a single pion or kaon, so most of the neutrons here are being produced through $(\gamma,n)$ processes and neutronic showers.  If we remake this figure, instead requiring, for example, $N_n \ge 3$, the results change significantly, with the fraction of neutron-producing events becoming much smaller.  Moreover, showers at high energy should always be cut~\cite{Li:2}, making it even more rare for relevant electromagnetic showers to have neutrons. These points mean that neutrons are extremely \textit{inefficient} at identifying electromagnetic showers, which is also good.

\begin{figure}[t]
\includegraphics[width=\columnwidth]{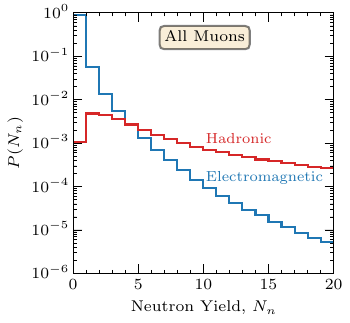}
\caption{Probability distributions of neutron yields of muons in Super-K, integrated over the muon spectrum.  These are normalized per vertical throughgoing muon.}
\label{fig:MuonsNeutrons}
\end{figure}

Figure~\ref{fig:MuonsNeutrons} shows the neutron yield distributions for the hadronic and electromagnetic showers that occur in Super-K when we integrate over the muon spectrum.  Because the muon rate is about 2~s$^{-1}$ and there are $\sim$$10^5$ seconds per day, a relative frequency of $10^{-5}$ corresponds to a rate of about two events per day, for example.  For electromagnetic showers, the yield distribution peaks at zero neutrons and then falls steeply, with an average of $N_n = 0.14$.  For hadronic showers, it is common to produce multiple neutrons, with the yield distribution falling slowly and extending up to thousands of neutrons (see Appendix~\ref{appendix:Yields}), with an average of $N_n = 19$. Those with $N_n = 0$ are mostly low-energy hadronic showers that either produce no neutrons or produce one neutron that undergoes an $(n,p)$ interaction. These points further illustrate our remarks about how Fig.~\ref{fig:HadronicNeutrons} and Fig.~\ref{fig:emNeutrons} would change if we changed the criterion from $N_n \ge 1$ to, say, $N_n \ge 3$.

\subsection{Isotopes are produced with neutrons}
\label{subsec:IsotopesNeutrons}

\begin{figure}[t]
\includegraphics[width=\columnwidth]{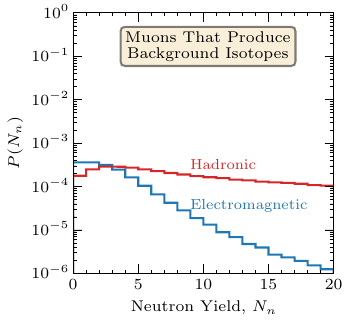}
\caption{The same as Fig.~\ref{fig:MuonsNeutrons}, but now only for muons that produce isotopes.}
\label{fig:IsotopeMuonsNeutrons}
\end{figure}

Figure~\ref{fig:IsotopeMuonsNeutrons} shows the neutron yield distributions for only showers that produce isotopes. For electromagnetic showers, the frequency drops dramatically relative to Fig.~\ref{fig:MuonsNeutrons} because most electromagnetic showers do not produce isotopes, while for hadronic showers, the frequency drops more modestly. At low $N_n$, the drops are especially large because showers with fewer neutrons are less efficient at producing isotopes. For the rare electromagnetic showers that produce isotopes, the average neutron yield is $N_n = 2.7$, which means neutrons can be used to tag isotope-producing showers even if they are electromagnetic.  At large $N_n$ for the hadronic case (for which the average is now $N_n = 60$), the fact that the distribution barely drops indicates the high probability of isotope production in such cases.  When we restrict to isotope-producing cases, electromagnetic showers are suppressed to a frequency comparable to that of hadronic showers.  Further, it is enough to cut on a small number of neutrons, making our technique sensitive to even small showers.  \textit{Together, these points mean that one simple cut can greatly reduce spallation backgrounds while incurring little deadtime.}

\begin{figure*}[t]
\includegraphics[width=\textwidth]{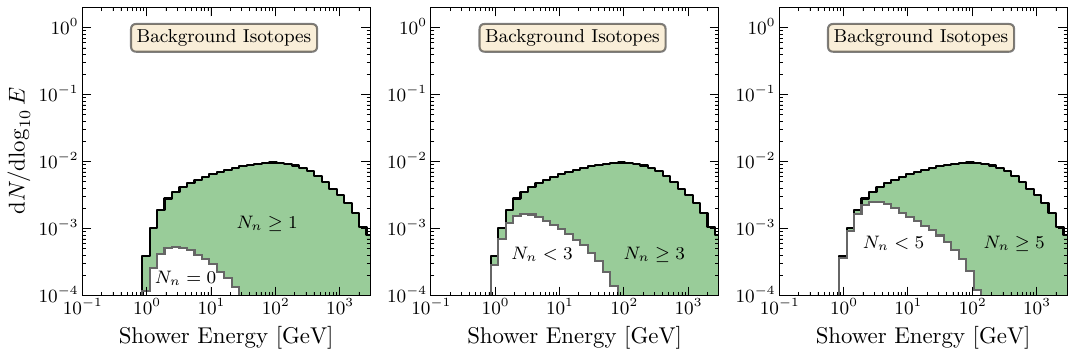}	
\caption{Parent shower energy distributions corresponding to the production of isotopes, normalized per vertical throughgoing muon.  We separate the distributions into components based on different neutron yield criteria. }
\label{fig:IsotopesNeutrons}
\end{figure*}

\begin{figure}[t]
\includegraphics[width=\columnwidth]{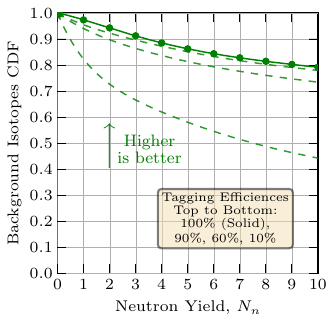}
\caption{Cumulative distribution for the fraction of isotopes produced with a neutron yield at or above a given value, for 100\%, 90\%, 60\%, 10\% neutron tagging efficiencies. Values are given in Table~\ref{table:CDF} in the Appendices.}
\label{fig:CumulativeIsotopes}
\end{figure}

\begin{figure}[t]
\includegraphics[width=\columnwidth]{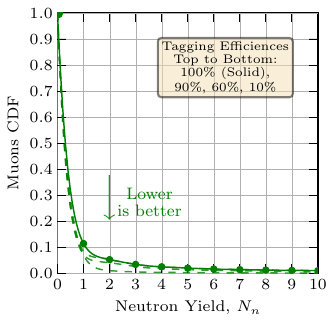}
\caption{Cumulative distribution for the fraction of
muons with a neutron yield at or above a given value, for 100\%, 90\%, 60\%, 10\% neutron tagging efficiencies. Values are given in Table~\ref{table:CDF} in the Appendices.}
\label{fig:CumulativeMuons}
\end{figure}

Figure~\ref{fig:IsotopesNeutrons} shows the energy spectra of the muon-induced showers that produce isotopes, weighted by the number of isotopes produced by each shower (which can be more than one for large showers).  The general shape can be explained as the product of the muon-induced hadronic shower spectrum in Fig.~\ref{fig:ShowersEnergySpectrum} with the increasing average isotope yields of hadronic showers, as shown in the Appendix~\ref{appendix:Yields}.  The energy spectrum of hadronic showers falls as $\sim$$E^{-0.5}$ and the isotope yield of hadronic showers rises as $\sim$$E^{0.9}$, so the energy spectrum of isotopes in Fig.~\ref{fig:IsotopesNeutrons} rises as $\sim$$E^{0.4}$ up until energies of $\sim$100~GeV, after which the spectrum falls again due to the drop in the muon spectrum.  At energies of $\sim$30~GeV, the average isotope yield per hadronic shower is $\sim$1, and above that energy multiple isotopes are often produced in the same hadronic shower. Cutting muons with large energy losses was suggested in Ref.~\cite{Li:2} and employed by Super-K with their multiple spallation cut~\cite{Super-Kamiokande:2021snn}. Fig.~\ref{fig:IsotopesNeutrons} shows that these large showers also produce many neutrons.

The lower the threshold on the neutron yield, the more efficient the cut will be in rejecting hadronic showers, and therefore in cutting isotopes.  How low in yield we can go depends on the neutron detection efficiency.  Even if this efficiency were perfect, we would miss the small fraction of hadronic showers with zero neutrons, either because the shower energy is low or because the neutrons undergo isotope-producing interactions.

Figure~\ref{fig:CumulativeIsotopes} --- one of our main results --- shows the fraction of isotopes produced with a neutron yield at or above a given value, assuming different neutron tagging efficiencies. In other words, it shows the fractions of the integral contributed by the green shaded regions in Fig.~\ref{fig:IsotopesNeutrons} for different $N_n$ thresholds. The $x$ axis here is the detected neutron yield, as opposed to the actual neutron yield. In the case of 100\% neutron tagging efficiency, these are the same, but for lower efficiencies, the detected neutron yield is calculated using Poisson statistics with an expectation value equal to the actual neutron yield multiplied by the tagging efficiency. 

Fig.~\ref{fig:CumulativeIsotopes} represents the fraction of isotopes that would be eliminated if all muons with certain neutron yields were rejected. For example, with 90\% tagging efficiency, more than 95\% of isotopes can be rejected by applying cuts to muons with $N_n \ge 1$. Even a higher and more conservative criterion of $N_n \ge 3$ would reject about 90\% of isotope backgrounds, reaching an efficiency comparable to that of current cuts, while affecting a much smaller fraction of muons, as we show below. For Super-K with pure water, where the neutron detection efficiency is $\sim$10\%, our method can still be beneficial. A threshold of $N_n \ge 1$ corresponds to an actual neutron yield of order $\sim$10, which can reject up to 80\% of isotopes with negligible deadtime.


\subsection{Neutron-producing muons are rare}
\label{subsec:NeutronsRare}

To estimate the deadtime that cuts on neutron-producing muons would cause, we need to calculate the fraction of muons that would be affected by those cuts.

Figure~\ref{fig:CumulativeMuons} --- another of our main results --- shows the fraction of muons with neutron yields at or above a given value, assuming different tagging efficiencies. These values represent the fraction of the integral contributed by the bins in Fig.~\ref{fig:MuonsNeutrons} above different $N_n$ thresholds. The sharp drop in the fraction of muons from $N_n=0$ to $N_n=1$ shows that about 89\% of muons that cross Super-K produce no neutrons at all. This number results from the facts that each through-going muon produces on average one shower (as defined above), $\sim$97\% of these muon-induced showers are electromagnetic (Fig.~\ref{fig:ShowersEnergySpectrum}), and $\sim$91\% of those electromagnetic showers do not produce neutrons (Fig.~\ref{fig:emNeutrons}).  Since the rate of all muons is $\sim$2~s$^{-1}$, the rate of neutron-producing muons is only $\sim$0.2~s$^{-1}$.

The results in Figs.~\ref{fig:CumulativeIsotopes} and \ref{fig:CumulativeMuons} show that increasing the threshold for the neutron yield would affect significantly fewer muons, while only slightly decreasing the background rejection rate. For example, the previously discussed $N_n \ge 3$ criterion rejects up to $90$\% of the background (Fig.~\ref{fig:CumulativeIsotopes}), while only affecting 3.5\% of the muons, corresponding to a rate of $\sim$0.07~s$^{-1}$.

With limited tagging efficiencies, these fractions are even smaller. In pure water ($\sim$10\% tagging efficiency), the contribution from fake neutrons was taken into account by adding their occurrence rate (0.044 per muon~\cite{Super-Kamiokande:2021snn}). In fact, the fraction of muons with one neutron or more ($\sim$5\%) is dominated by this contribution from fake neutrons. However, their overall fraction of all muons is still very small.

\subsection{Recommendations for implementation}

The primary point of our new methods is that much harder cuts can be applied to a much smaller fraction of muons.  Throughout this paper, we focus on the physics of neutron production with isotopes, making various simplifications on other aspects (e.g., assuming vertical throughgoing muons).  To estimate the \textit{relative} impact of our new methods, we use simple cuts for the before and after assessments.  For the \textit{absolute} impact, it will be important to use the likelihood methods developed by Super-K (and used by us in Ref.~\cite{Li:3}).  Because we make conservative choices, we expect the absolute impact to be even better than the relative impact.

\begin{figure}[t]
\includegraphics[width=\columnwidth]{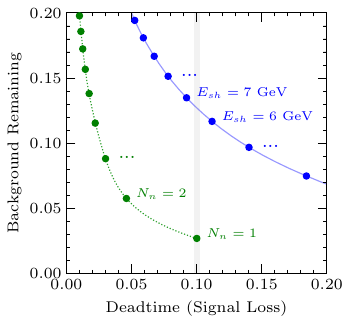}
\caption{Deadtime and efficiency comparison for cylinder cuts based on shower energy (blue solid line) and neutron yields (dotted green line). The blue and green points represent different thresholds for $E_{\text{sh}}$ and $N_n$ with increments of 1~GeV and 1~neutron, respectively. The gray band highlights the efficiency improvement at $\sim$10\% deadtime.}
\label{fig:Deadtime}
\end{figure}

To assess the relative impact of our new methods, we approximate Super-K's likelihood cut with a collection of hard cuts that are applied to a small fraction of muons.  These start with a cylinder cut, characterized by the time delay ($t$) and the transverse distance to the muon track ($L_{\text{trans}}$).  We choose 35~s for the time delay, which is about three time longer than the mean lifetime of $^{16}$N, an isotope with a large production rate, decay energy, and lifetime. We set the cylinder radius to be 3~m, large enough to capture nearly all activity (see Appendix~\ref{appendix:PhysicalDistributions} for details).  We shorten the length of the cylinder by exploiting the correlation in the longitudinal separation between the isotope position and the shower peak ($L_{\text{long}}$); for this, we use 10~m (see Appendix~\ref{appendix:PhysicalDistributions}).  Last, as detailed below, we add a cut on either the radiation energy loss (shower energy, $E_{\text{sh}}$, which Super-K calls $Q_{res}$) or the number of neutrons.  

In this simple treatment, all data that falls inside the individual hard cuts would simply be discarded. We then calculate the remaining fraction of backgrounds. The deadtime is estimated as the fraction of the lost exposure due to these cylinder cuts (neglecting overlaps). A more accurate way of estimating the deadtime would be to generate a random sample of events uniformly distributed in space and time, calculating the fraction of this sample that is affected by the cuts. The latter method gives slightly smaller deadtimes since it takes into account overlaps of the muons, but here we use the former as a simple and conservative estimate.

Figure~\ref{fig:Deadtime} (\textit{before case}) shows the deadtime and background-rejection efficiency for cylinder cuts where we vary the threshold for $E_{\text{sh}}$, meaning that showers above a certain energy are always rejected.  These cuts can approximate (for a cut-off between 6--7~GeV) the performance of the current likelihood used by Super-K, which rejects $\sim$90\% of the backgrounds with $\sim$10\% deadtime.

Figure~\ref{fig:Deadtime} (\textit{after case}) also shows the deadtime and background-rejection efficiency for similar cylinder cuts when we vary the threshold for $N_n$.  At $\sim$10\% deadtime, applying cylinder cuts to muons based on their neutron yields can reduce the remaining background by a factor of $\sim$4 compared to cuts based on their shower energy.  \textit{One of our key results is that neutron yields are far better indicator of a shower's likelihood to produce isotopes than its energy.}  

In a fuller treatment, we would take into account the correlations of the cut  variables, the overlap of muon tracks in assessing the deadtime, and new cuts on the muons that are not subjected to hard cuts.  All of these would act in the sense of giving better results.


\section{Conclusions}
\label{sec:Conclusions}

Progress in MeV-range neutrino astronomy is limited by detector backgrounds.  One of the most significant is delayed beta decays of unstable isotopes produced by cosmic-ray muons.  In Super-K, which is our focus, these spallation backgrounds are significant even after cuts developed over decades of experimental work.  For solar neutrinos, spallation backgrounds degrade Super-K's sensitivity to the day-night effect, the shape of the $^8$B spectrum, and the search for the $hep$ flux~\cite{Gann:2021ndb}.  For the DSNB, spallation backgrounds and those due to atmospheric neutrinos~\cite{Beacom:2010kk, Super-Kamiokande:2021jaq, Zhou:2023mou} degrade Super-K's sensitivity at lower and higher energies, respectively.  These difficulties are also important for other neutrino detectors.

Here we show a new way forward, taking advantage of the addition of gadolinium to Super-K to enable neutron detection and our extensive earlier work on spallation~\cite{Galbiati:2005ft, Li:1, Li:2, Li:3, Zhu:2018rwc}.  We find --- based on the results of careful simulations and key theoretical insights --- that isotope production is almost always accompanied by neutron production and that muons rarely produce either.  This means that the relevant muons for isotope production can be identified through neutron detection.  By applying stronger cuts to a smaller fraction of muons, we can simultaneously improve cut efficiency and deadtime.  \textit{Our results will increase the effective depth of detectors.}  In more detail, the main benefits are as follows:
\begin{enumerate}
    
\item \textbf{For Super-K with gadolinium.}  Because the neutron detection efficiency is high, it should be possible to identify relevant muons through the detection of even a single neutron.  This should reduce spallation backgrounds by a factor of at least four with a low deadtime.  In combination with other spallation-reduction techniques, it should be possible to nearly eliminate spallation backgrounds.  This would apply to data taken since August of 2020 (SK-Gd)~\cite{Super-Kamiokande:2021the}.

\item \textbf{For Super-K without gadolinium.}  With the lower neutron-detection efficiency of pure water, it would still be possible to tag muons that produce $\sim$10 neutrons with at least one detectable neutron capture.  Even this should be a helpful new cut, especially when combined with other spallation-reduction techniques.  This could be applied to reanalyze substantial earlier data (SK-IV, for which adequate low-level information was saved)~\cite{Super-Kamiokande:2023jbt}.

\item \textbf{For Hyper-K.}  Compared to Super-K (at a depth of 1000~m of rock), Hyper-K will be at a shallower depth of 600~m, where the muon flux is about five times larger~\cite{Hyper-Kamiokande:2018ofw}.  With added gadolinium, Hyper-K could greatly reduce spallation backgrounds and thus take full advantage of its large size, leading to breakthroughs in solar and DSNB science.

\end{enumerate}
In addition, there are bonus benefits:
\begin{enumerate}
    
\item \textbf{For JUNO and other detectors.}  JUNO will be at a depth of 700~m, where the muon flux is about two times larger than at Super-K~\cite{JUNO:2015sjr}.  While there will be adjustments needed in the details of our predictions, the basic ideas of our approach will work for JUNO and other scintillator detectors, as well as for other detector types.

\item \textbf{As a new tool for supernova calibration.}  A Milky Way supernova is expected to create $\sim$$10^4$ events in Super-K.  We find that large bursts of neutron captures (and isotope production) happen frequently enough in Super-K that they could be used to test the supernova response.  See Appendix~\ref{appendix:Yields} for details.

\end{enumerate}


\bigskip
\section*{Acknowledgements}

We are grateful for helpful discussions with Sonia El-Hedri, Mark Vagins, and especially Michael Smy. The work of O.~N. and J.~F.~B. was primarily supported by National Science Foundation Grant No.\ PHY-2310018 (with preliminary work supported by National Science Foundation Grant No.\ PHY-2012955).

\section*{Data Availability}

The data associated with the figures in this paper are available via Zenodo~\cite{ZENODO}.


\clearpage
\appendix

\centerline{\Large {\bf Appendices}} 

\vspace{0.75cm}

Here we provide further details --- including some new results --- on yields, spatial distributions, and comparisons to spallation in scintillator detectors. For neutrons, we record their positions at capture (while the presence of gadolinium shortens the capture time, it hardly changes the capture positions).  For isotopes, we record their positions at production (we neglect motion caused by convective currents, as Super-K data suggests that these are small~\cite{Super-Kamiokande:2016yck, Super-Kamiokande:2021the, Super-Kamiokande:2024kcb}).  For electrons, we record the positions where their energy falls below 100~keV, the default transport cutoff energy in \texttt{FLUKA}.


\section{Final-state yields from showers}
\label{appendix:Yields}

\begin{figure}[b]
    \includegraphics[width=\columnwidth]{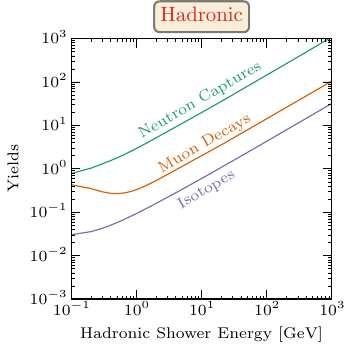}
    \newline \vspace{0.1cm}
    \includegraphics[width=\columnwidth]{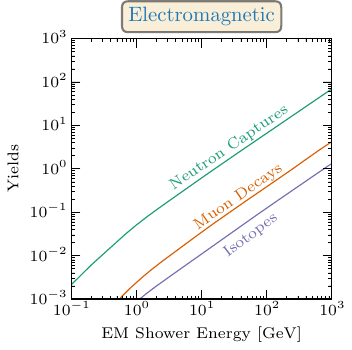}
\caption{Numbers of neutron captures, muon decays, and isotopes left after hadronic and electromagnetic showers as a function of their energy. The differences between the hadronic and electromagnetic showers are large.}
\label{fig:Yields}
\end{figure}

As a shower develops, the number of particles in it rises and then falls.  Here we provide further details on the yields of the residual particles remaining after a shower has ended.  Because typical showers in Super-K have lengths of up to several meters, the duration of the shower is at most a few tens of ns, with the path lengths and timescales of the last shower particles being even less.  \textit{The residuals we consider produce detectable signals on much longer timescales.}

The residuals may include charged pions and muons, neutrons, and isotopes, eventually all at rest (more accurately, thermal energies).  Negative pions and muons undergo atomic and then nuclear capture, transforming protons into neutrons and often breaking nuclei~\cite{Ponomarev:1973ya, Measday:2001yr}.  Positive pions decay into positive muons with a mean life of 26~ns, which then decay into positrons with a mean life of 2.2~$\mu$s.  Neutrons undergo capture in tens of $\mu$s, depending on the gadolinium concentration.  And isotopes decay with mean life values up to several seconds.

Figure~\ref{fig:Yields} shows the neutron, muon, and isotope yields in hadronic and electromagnetic showers as a function of the shower energy.  The main feature is that the yields are approximately one order of magnitude lower in the electromagnetic case.  In the hadronic case at the lowest energies, the results are significantly affected by the fact that half of the pions we inject are negative.  In scintillator detectors, the muon and neutron yields are comparable, but the isotope yield is much higher, principally due to $^{11}$C (see Apeendix.~\ref{appendix:Scintillator}).

\begin{figure}[h]
\centering
\includegraphics[width=\columnwidth]{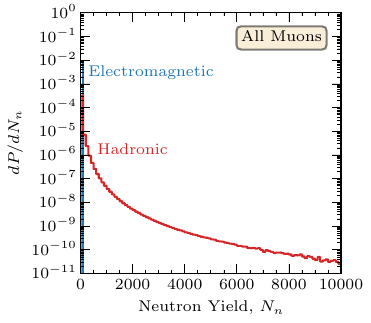}
\caption{Neutron yield distribution for muons at Super-K, similar to Fig.~\ref{fig:MuonsNeutrons}, but extending to larger neutron yields and using bin widths of of $dN_n = 100$ instead of 1.}
\label{fig:LargeNeutronYields}
\end{figure}

Figure~\ref{fig:LargeNeutronYields} shows the neutron yield distributions for muon-induced hadronic and electromagnetic showers at Super-K.  Here we extend to much larger yields than in Fig.~\ref{fig:MuonsNeutrons}, so we take the continuum limit, showing $dP(N_n)/dN_n$, with $dN_n = 100$.  The two figures would agree if the yield distributions were near flat; they are not, so this figure should be viewed as approximate.  The key point is that muons can often produce very large numbers of neutrons.  Near $N_n = 1000$, for example, $dP(N_n)/dN_n \sim 10^{-7}$, so the probability of observing a number of neutrons in the range 1000--1099 is $\sim$$10^{-5}$.  Because the muon rate is about 2~s$^{-1}$ (and there are $\sim$$10^5$ seconds per day), that means that yields of neutrons this large happen daily.  As another example, neutron yields in the range 8000--8099 happen on a yearly basis.

\begin{figure}[h]
\centering
\includegraphics[width=\columnwidth]{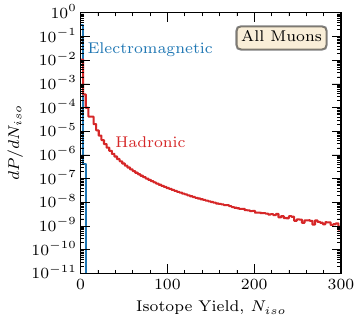}
\caption{Isotope yield distribution for muons at Super-K, with bin widths of $dN_{iso} = 3$.}
\label{fig:LargeIsotopeYields}
\end{figure}

Figure~\ref{fig:LargeIsotopeYields} shows the isotope yield distributions for muons at Super-K. This is related to Fig.~\ref{fig:LargeNeutronYields} by a change of variables from $N_n$ to $N_{iso}$. The neutron yields of hadronic showers are $\sim$34 times larger than the isotope yields (see Fig.~\ref{fig:Yields}), so here the bin widths are changed to be $dN_{iso} = 3$. The points we mentioned regarding the frequency of large neutron yields apply to isotope yields as well. For example, the daily muons with $\sim$1000 neutrons also produce $\sim$30 background isotope events, and the yearly muons with $\sim$8000 neutrons produce $\sim$240 isotopes. This large number of events can be utilized to test and calibrate the detector response to a supernova event.

\section{Spatial profiles}
\label{appendix:PhysicalDistributions}

The spatial profiles of muon-induced showers are needed to determine the cuts necessary to fully contain these showers. Here we calculate the average longitudinal and lateral profiles of typical muon-induced showers.  For electromagnetic showers, the average and individual longitudinal profiles --- and the physics behind them --- are discussed in Ref.~\cite{Li:2}.  For hadronic showers, the average longitudinal profiles are similar to those for electromagnetic showers, with some differences due to the pion multiplicity.  In addition for hadronic showers, the individual longitudinal profiles have larger fluctuations than for electromagnetic showers~\cite{Li:2}.

\begin{figure}[h]
\centering
\includegraphics[width=0.49\textwidth]{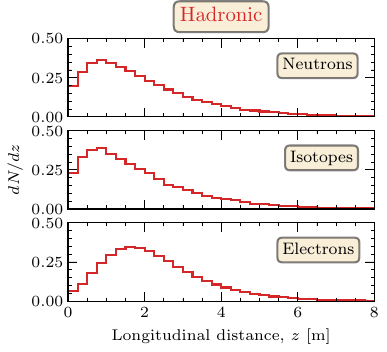}
\includegraphics[width=0.49\textwidth]{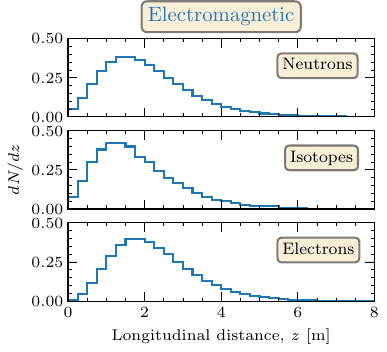}
\caption{Average longitudinal profiles of neutron captures, isotope production locations, and electron stopping points in 10-GeV hadronic and electromagnetic showers. The differences between hadronic and electromagnetic showers are small.}
\label{fig:Longitudinal}
\end{figure}

Here we provide more details, focusing on 10-GeV showers as being representative of moderate-energy showers, defined as being electromagnetic or hadronic based on the injected particles, as in Sec.~\ref{sec:NeutronProductioninShowers}.  For the average results, we sample over many individual showers.  For different shower energies (not shown), the average profiles are similar.  As the shower energy increases, the fluctuations decrease and the longitudinal shower extent slowly increases~\cite{Li:2}. The lateral extent of the showers varies very little as the energy increases.

Figure~\ref{fig:Longitudinal} shows the average longitudinal profiles of neutron captures, isotope production, and electron stopping points in 10-GeV electromagnetic and hadronic showers.  Both have a longitudinal extent of $\sim$6~m, and the shapes of their longitudinal profiles have small differences. One of the main differences is that the neutron and isotope profiles in hadronic showers peak at a smaller distance. This is because the initial interactions of the injected pions can produce neutrons and isotopes.

\begin{figure}[h]
    \includegraphics[width=\columnwidth]{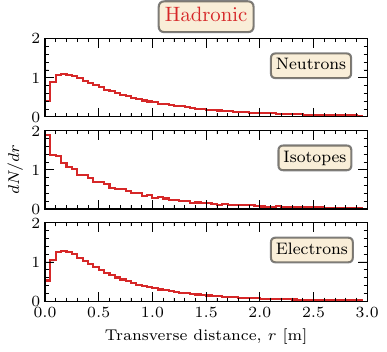}
    \includegraphics[width=\columnwidth]{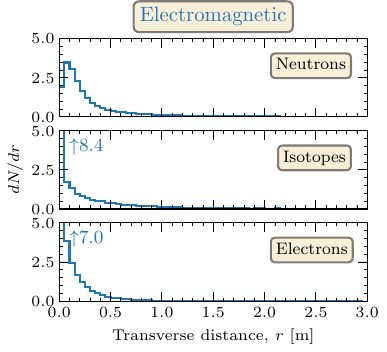}
\caption{Average lateral profiles of neutron captures, relevant isotope production locations, and electron stopping points in 10-GeV hadronic and electromagnetic showers. The differences between hadronic and electromagnetic showers are large.}
\label{fig:Lateral}
\end{figure}

Figure~\ref{fig:Lateral} shows the average lateral profiles of neutron captures, isotope production, and electron stopping points around the tracks of 10-GeV electromagnetic and hadronic showers. The area factor $2\pi r dr$ is included. Compared to the longitudinal profiles, the lateral profiles of electromagnetic showers and hadronic showers have large differences. In earlier work, we have shown that the lateral profiles of different isotopes are quite similar~\cite{Li:1}.

For electromagnetic showers, the lateral extent is characterized by the Molière radius ($\sim$10 cm in water). The majority of the shower energy is deposited within 0.5~m around the shower track, as seen in the bottom panel of Fig.~\ref{fig:Lateral}. At 10~GeV, electromagnetic showers have an average neutron yield of $\sim$0.6. Neutrons at energies above tens of MeV can travel far away from the track because they undergo many scattering interactions before capturing, unlike thermal neutrons, but most of them still capture within $\sim$1.5~m from the shower. Isotope production in electromagnetic showers is mainly caused by these neutrons, so the isotopes are also contained within $\sim$1.5~m, but there is a larger peak at close distances to the shower because of isotopes produced by gamma rays directly or by neutrons that interact quickly.

On the other hand, hadronic showers have a larger lateral extent because of the much larger mean free paths for pions and fast neutrons.  In this case, one needs a radius of about $\sim$3~m to contain all of the particles.  (Even for low-energy hadronic showers, the lateral extent remains large.)  The distribution of isotope production locations peaks closer to the shower track because of isotopes produced directly by the injected pions.


\section{Comparisons to scintillator detectors}
\label{appendix:Scintillator}

Our proposal to reduce spallation backgrounds by neutron tagging of showers is significantly different from a technique first developed by Borexino and used for other scintillator detectors~\cite{Galbiati:2004wx, Borexino:2011cjz, Borexino:2021pyz, KamLAND-Zen:2023spw}.  This technique is designed to reduce backgrounds due to $^{11}$C, an important spallation product. This isotope can be produced by a gamma ray from a muon-induced electromagnetic shower initiating a $(\gamma,n)$ interaction.  For possible $^{11}$C beta-decay events, one evaluates a three-fold coincidence between the possible decay event and the distances to the muon and the nearest neutron capture. In contrast, our proposal is based on \textit{the presence of neutrons due to hadronic showers, preceding isotope production}. Recently, Borexino used a neutron-multiplicity technique to improve on the three-fold coincidence veto for $^{11}$C~\cite{Borexino:2021pyz}.  In related work, KamLAND-Zen utilized neutron multiplicities for spallation studies on xenon~\cite{KamLAND-Zen:2023spw} Our approach is significantly more general than these --- e.g., in that it applies to many isotopes --- and is supported by the extensive theoretical work above.

In more detail, there are some key similarities and differences between spallation in scintillator and water.  The neutron yields per muon are comparable.  And the production rate of $^{11}$C in scintillator is comparable to that of $^{15}$O in water, which is about $\sim$0.1 of the neutron yield, both dominated by $(\gamma,n)$ processes.  However, while $^{11}$C is an important background in scintillator, $^{15}$O is not one in water because its beta-decay energy is so low.  All other isotopes in both scintillator and water detectors --- which are produced at much lower levels --- are primarily caused by hadronic processes.

Due to these points, our results would have some different features in scintillator detectors. In Fig.~\ref{fig:IsotopeMuonsNeutrons}, the $N_n = 1$ bin would have a much larger fraction of the isotope-producing muons because of the very frequent electromagnetic showers that produce $^{11}$C through $(\gamma,n)$, which yields one neutron. This means that the efficiency would decrease rapidly if the threshold were increased from $N_n = 1$ to any larger value.  In Fig.~\ref{fig:CumulativeIsotopes}, the cumulative fraction of isotopes would drop sharply from $\sim$97\% at $N_n = 1$ to about $\sim$70\% at $N_n =2$ in scintillator detectors (compared to $\sim$95\% in water detectors). Therefore, this technique would require a perfect neutron capture detection efficiency (for $N_n = 1$) to achieve a high rejection rate, but in water detectors neutron tagging can still work well even with a limited neutron detection efficiency. 


\onecolumngrid
\begin{center}

\section{Cumulative fractions}
\label{appendix:fractions}

Table~\ref{table:CDF} shows the values of the cumulative distribution functions for background isotopes and muons in Figs.~\ref{fig:CumulativeIsotopes} and~\ref{fig:CumulativeMuons}.

\begin{table}[h]
\centering
\renewcommand{\arraystretch}{1.3}
\begin{tabular}{c|cccc|cccc}
    \toprule
    \multicolumn{1}{c}{ } & \multicolumn{4}{c}{\textbf{Cumulative Fraction of Isotopes}} & \multicolumn{4}{c}{\textbf{Cumulative Fraction of Muons}} \\
    \cmidrule(lr){2-5} \cmidrule(lr){6-9}
    \multicolumn{1}{c}{ }& \textbf{100\%}  &\textbf{90\%}  & \textbf{60\%}  & \multicolumn{1}{c}{ \textbf{10\%}}  & \textbf{100\%}  & \textbf{90\%}  & \textbf{60\%}  & \textbf{10\%}  \\
    \midrule
    \boldmath$N_n \ge 0$ & 1.0000 & 1.0000 & 1.0000 & 1.0000 & 1.0000 & 1.0000 & 1.0000 & 1.0000 \\
    \boldmath$N_n \ge 1$ & 0.9732 & 0.9529 & 0.9388 & 0.8206 & 0.1145 & 0.0856 & 0.0728 & 0.0699 \\
    \boldmath$N_n \ge 2$ & 0.9425 & 0.9238 & 0.8983 & 0.7285 & 0.0527 & 0.0551 & 0.0411 & 0.0104 \\
    \boldmath$N_n \ge 3$ & 0.9119 & 0.8958 & 0.8637 & 0.6666 & 0.0344 & 0.0362 & 0.0256 & 0.0055 \\
    \boldmath$N_n \ge 4$ & 0.8846 & 0.8714 & 0.8356 & 0.6186 & 0.0253 & 0.0260 & 0.0182 & 0.0039 \\
    \boldmath$N_n \ge 5$ & 0.8619 & 0.8504 & 0.8124 & 0.5789 & 0.0200 & 0.0202 & 0.0142 & 0.0030 \\
    \boldmath$N_n \ge 6$ & 0.8433 & 0.8324 & 0.7929 & 0.5448 & 0.0167 & 0.0166 & 0.0117 & 0.0024 \\
    \boldmath$N_n \ge 7$ & 0.8276 & 0.8168 & 0.7758 & 0.5149 & 0.0144 & 0.0142 & 0.0101 & 0.0020 \\
    \boldmath$N_n \ge 8$ & 0.8142 & 0.8030 & 0.7606 & 0.4882 & 0.0128 & 0.0124 & 0.0089 & 0.0017 \\
    \boldmath$N_n \ge 9$ & 0.8022 & 0.7906 & 0.7467 & 0.4643 & 0.0115 & 0.0111 & 0.0080 & 0.0015 \\
    \boldmath$N_n \ge 10$ & 0.7915 & 0.7793 & 0.7339 & 0.4425 & 0.0105 & 0.0101 & 0.0072 & 0.0013 \\
    \boldmath$N_n \ge 11$ & 0.7814 & 0.7688 & 0.7219 & 0.4226 & 0.0097 & 0.0093 & 0.0066 & 0.0012 \\
    \boldmath$N_n \ge 12$ & 0.7720 & 0.7591 & 0.7105 & 0.4043 & 0.0091 & 0.0086 & 0.0061 & 0.0010 \\
    \boldmath$N_n \ge 13$ & 0.7632 & 0.7498 & 0.6998 & 0.3874 & 0.0085 & 0.0080 & 0.0057 & 0.0009 \\
    \boldmath$N_n \ge 14$ & 0.7547 & 0.7410 & 0.6896 & 0.3717 & 0.0080 & 0.0075 & 0.0053 & 0.0009 \\
    \boldmath$N_n \ge 15$ & 0.7467 & 0.7326 & 0.6798 & 0.3570 & 0.0075 & 0.0071 & 0.0050 & 0.0008 \\
    \boldmath$N_n \ge 16$ & 0.7390 & 0.7246 & 0.6705 & 0.3433 & 0.0071 & 0.0067 & 0.0047 & 0.0007 \\
    \boldmath$N_n \ge 17$ & 0.7314 & 0.7169 & 0.6615 & 0.3305 & 0.0068 & 0.0063 & 0.0045 & 0.0006 \\
    \boldmath$N_n \ge 18$ & 0.7241 & 0.7094 & 0.6528 & 0.3184 & 0.0065 & 0.0060 & 0.0042 & 0.0006 \\
    \boldmath$N_n \ge 19$ & 0.7171 & 0.7022 & 0.6445 & 0.3071 & 0.0062 & 0.0057 & 0.0040 & 0.0005 \\
    \boldmath$N_n \ge 20$ & 0.7103 & 0.6953 & 0.6365 & 0.2965 & 0.0059 & 0.0055 & 0.0038 & 0.0005 \\
    \bottomrule
    \end{tabular}
\caption{Cumulative fraction of isotopes and muons as a function of the neutron yield for different tagging efficiencies. }
\label{table:CDF}
\end{table}
\end{center}
\twocolumngrid

\clearpage
\bibliography{spallation}

\end{document}